\documentclass[aip,apl,12pt, preprint]{revtex4-1}

\usepackage{graphicx}
\usepackage{amsmath, amsthm, amssymb}

\usepackage{siunitx} 
\begin{document}

\preprint{}

\title[]{Integrated waveguides and deterministically positioned nitrogen vacancy centers in diamond created by femtosecond laser writing} 




\author{J. P. HADDEN}
\altaffiliation{These authors contributed equally to this publication}
\affiliation{Institute for Quantum Science and Technology, University of Calgary, Calgary, T2N 1N4, Canada.}

\author{V. BHARADWAJ}
\altaffiliation{These authors contributed equally to this publication}
\affiliation{Istituto di Fotonica e Nanotecnologie-Consiglio Nazionale delle Ricerche (IFN-CNR) and Dipartimento di Fisica - Politecnico di Milano, Piazza Leonardo da Vinci 32, Milano, 20133, Italy}

\author{B. SOTILLO}
\affiliation{Istituto di Fotonica e Nanotecnologie-Consiglio Nazionale delle Ricerche (IFN-CNR) and Dipartimento di Fisica - Politecnico di Milano, Piazza Leonardo da Vinci 32, Milano, 20133, Italy}

\author{S. RAMPINI}
\affiliation{Istituto di Fotonica e Nanotecnologie-Consiglio Nazionale delle Ricerche (IFN-CNR) and Dipartimento di Fisica - Politecnico di Milano, Piazza Leonardo da Vinci 32, Milano, 20133, Italy}

\author{R. OSELLAME}
\affiliation{Istituto di Fotonica e Nanotecnologie-Consiglio Nazionale delle Ricerche (IFN-CNR) and Dipartimento di Fisica - Politecnico di Milano, Piazza Leonardo da Vinci 32, Milano, 20133, Italy}

\author{J. WITMER}
\altaffiliation{Currently at Ginzton Laboratory, Stanford University, Stanford, CA 94305, USA}
\affiliation{Institute for Quantum Science and Technology, University of Calgary, Calgary, T2N 1N4, Canada.}

\author{H. JAYAKUMAR}
\altaffiliation{Currently at Department of Physics, City University of New York (CUNY)-City College of New York, New York, NY 10031, USA}
\affiliation{Institute for Quantum Science and Technology, University of Calgary, Calgary, T2N 1N4, Canada.}

\author{T. T. FERNANDEZ}
\affiliation{Istituto di Fotonica e Nanotecnologie-Consiglio Nazionale delle Ricerche (IFN-CNR) and Dipartimento di Fisica - Politecnico di Milano, Piazza Leonardo da Vinci 32, Milano, 20133, Italy}

\author{A. CHIAPPINI}
\affiliation{Istituto di Fotonica e Nanotecnologie (IFN-CNR), CSMFO and FBK-CMM, Via alla Cascata, 56/C, Trento, 38123, Italy}

\author{C. ARMELLINI}
\affiliation{Istituto di Fotonica e Nanotecnologie (IFN-CNR), CSMFO and FBK-CMM, Via alla Cascata, 56/C, Trento, 38123, Italy}

\author{M. FERRARI}
\affiliation{Istituto di Fotonica e Nanotecnologie (IFN-CNR), CSMFO and FBK-CMM, Via alla Cascata, 56/C, Trento, 38123, Italy}

\author{R. RAMPONI}
\affiliation{Istituto di Fotonica e Nanotecnologie-Consiglio Nazionale delle Ricerche (IFN-CNR) and Dipartimento di Fisica - Politecnico di Milano, Piazza Leonardo da Vinci 32, Milano, 20133, Italy}

\author{P. E. BARCLAY}
\affiliation{Institute for Quantum Science and Technology, University of Calgary, Calgary, T2N 1N4, Canada.}

\author{S. M. EATON}
\affiliation{Istituto di Fotonica e Nanotecnologie-Consiglio Nazionale delle Ricerche (IFN-CNR) and Dipartimento di Fisica - Politecnico di Milano, Piazza Leonardo da Vinci 32, Milano, 20133, Italy}

\collaboration{Corresponding author: jpe.hadden@gmail.com}
\noaffiliation


\date{\today}

\begin{abstract}
Diamond's nitrogen vacancy (NV) center is an optically active defect with long spin coherence times, showing great potential for both efficient nanoscale magnetometry and quantum information processing schemes. Recently, both the formation of buried 3D optical waveguides and high quality single NVs in diamond were demonstrated using the versatile femtosecond laser-writing technique. However, until now, combining these technologies has been an outstanding challenge. In this work, we fabricate laser written photonic waveguides in quantum grade diamond which are aligned to within micron resolution to single laser-written NVs, enabling an integrated platform providing deterministically positioned waveguide-coupled NVs. This fabrication technology opens the way towards on-chip optical routing of single photons between NVs and optically integrated spin-based sensing.
\end{abstract}


\maketitle




%
%

%

Diamond's negatively charged nitrogen-vacancy (NV) color center is a point defect consisting of a nearest-neighbor subsitutional nitrogen atom and a lattice vacancy. These optically active defects have long room temperature spin coherence times, making them promising for nanoscale magnetic field sensing \cite{Taylor2008} and for quantum information processing systems \cite{Hensen2015}. An integrated photonics platform in diamond would be beneficial for quantum computing based on optically entangled NVs interacting via optical waveguides, as well as NV-based optically detected spin sensing application, due to the integration and stability provided by monolithic optical waveguides \cite{Schroder2016}.

 Waveguides and other planar photonic structures in diamond have previously been demonstrated within membranes produced using ion implantation assisted lift off methods \cite{Olivero2005}, plasma etching \cite{Faraon2011a}, or with under cutting performed through focused ion beam milling\cite{Castelletto2011}, angled reactive ion etching\cite{Burek2014a}, or quasi-isotropic etching\cite{Khanaliloo2015}.  However these techniques are restricted to 2D geometries, and up to now device lengths of up to \SI{\sim 100}{\micro\meter}.  Recently, the versatile femtosecond laser microfabrication method was used to demonstrate buried 3D optical waveguides in diamond \cite{Sotillo2016,Courvoisier2016}, a key step towards a diamond photonics platform. However, for practical implementation of quantum information systems and high spatial resolution magnetometry devices, deterministic placement of NVs is required. The most widely used method to place NV centers in diamond relies on ion implantation followed by annealing \cite{Orwa2011,Rabeau2006}. This method offers submicron spatial accuracy and significant progress has been made in the development of annealing processes to repair unwanted damage to the lattice which can degrade the NV centers properties, especially for shallow centers \cite{Naydenov2010, Yamamoto2013}. However the increased energy required for deeper NV creation causes even more damage \cite{Pezzagna2010}, thus techniques which place NV centers in buried structures with minimal damage are preferable.

In 2013, it was shown that NV centers could be produced in optical grade diamond by exploiting the electron plasma created by femtosecond laser pulses focused just above the diamond surface \cite{Liu2013}. This technique however was limited to near-surface placement of NVs and caused ablative damage of the diamond surface. Recently, Chen \textit{et al.} demonstrated a significant improvement, with on-demand and high quality single NVs written in the bulk of quantum grade diamond (nitrogen concentration of \SI{<5}{ppb}) using femtosecond laser writing with a spatial light modulator (SLM) to correct for aberrations followed by high temperature annealing \cite{Chen2016, Sotillo2017}. 

In this Letter, we apply femtosecond laser writing to inscribe buried optical waveguides and to create single NVs within quantum grade diamond. Crucially, we find that the optical waveguide remains even after the high temperature sample annealing required to form single NVs. Since the same laser and positioning system are used to form both the waveguide and NVs, they may be aligned to within micron resolution. We further demonstrate waveguide excitation and collection of light from laser-formed single NVs. Such waveguide coupled NVs could open the door to more sophisticated quantum photonic networks in diamond, exploiting optically linked single NVs for single photon sources or solid state qubits \cite{Kimble2008,Shi2016,Sipahigil2016}. In higher nitrogen content diamond, laser writing of high density NV ensembles within waveguides could enable robust excitation and collection of the fluorescence signal for spin based sensing \cite{Gazzano2016,Clevenson2015,Chen2017}.

Waveguides were written in quantum grade diamond using a regeneratively amplified Yb:KGW system with \SI{230}{\femto\second} pulse duration, \SI{515}{\nano\meter} wavelength and \SI{500}{\kilo\hertz} repetition rate, focussed with a 1.25 NA objective\cite{SupportingInfo}.  Figure 1 shows an overhead schematic (a) and optical microscope image (b) of the waveguide-NV optical device. Because focused femtosecond laser pulses yield amorphization and graphitization in crystalline diamond, the type II geometry \cite{Burghoff2006} was adopted, where two closely spaced laser-written modification lines separated by \SI{13}{\micro\meter} provide optical confinement.  The confinement mechanism has been shown to be reduction of the refractive index within the modification lines which provides lateral confinement, with the introduction of stress within the waveguide itself which provides polarization dependent vertical confinement\cite{Sotillo2017a}.  Modification lines spanning the \SI{2}{\milli\meter} sample were written with \SI{60}{\nano\joule} pulse energy and \SI{0.5}{\milli\meter/\second} scan speed leading to single mode guiding in the visible.  Multiple test devices were written to calibrate the depth and pulse energy of NV formation, however here we focus on a single waveguide with NV centers created near the waveguide depth. The mode field diameter (MFD) for the nearly circular TM mode was  \SI{9.5}{\micro\meter} at \SI{635}{\nano\meter} wavelength (Fig. 1(c)) and the insertion loss was \SI{5.4}{\decibel}. Accounting for the coupling loss and the Fresnel loss due to the refractive index mismatch, we infer a propagation loss of \SI{4.2}{\decibel/\centi\meter}.  These particular optical waveguides are linear, therefore the achievable bend loss has not been measured, however Courvoisier et al. showed a 3 dB bend loss for 25 mm bend radius using a similar laser writing method \cite{Courvoisier2016}. Future studies will seek to form curved waveguide structures such as directional couplers, and to characterize and optimize their bend loss.

Using the same femtosecond laser setup, single-pulse exposures were inscribed within the optical waveguide to induce vacancies in the diamond lattice.  Five identical single-pulse exposures (\SI{28}{\nano\joule}) separated by \SI{20}{\micro\meter} were written in each device in order to study reproducibility in forming single NV centers within the waveguiding region.  Markers visible in our imaging system were inscribed outside of the waveguide to locate the single pulse exposures, which are not visible to wide-field illuminated optical microscopy.  The diamond sample was subsequently annealed at \SI{1000}{\celsius} for 3 hours in a nitrogen atmosphere in order to form NV centers and anneal out other detrimental vacancy complexes\cite{Rabeau2006,Naydenov2010, Yamamoto2013}.  Crucially for the NV-waveguide device targeted in this work, we found that the mode profile and insertion loss of the annealed waveguides in diamond were unaltered.  For more details on device fabrication and annealing see Supplementary Material \cite{SupportingInfo}.

\begin{figure}
\centering
\includegraphics[width=\linewidth]{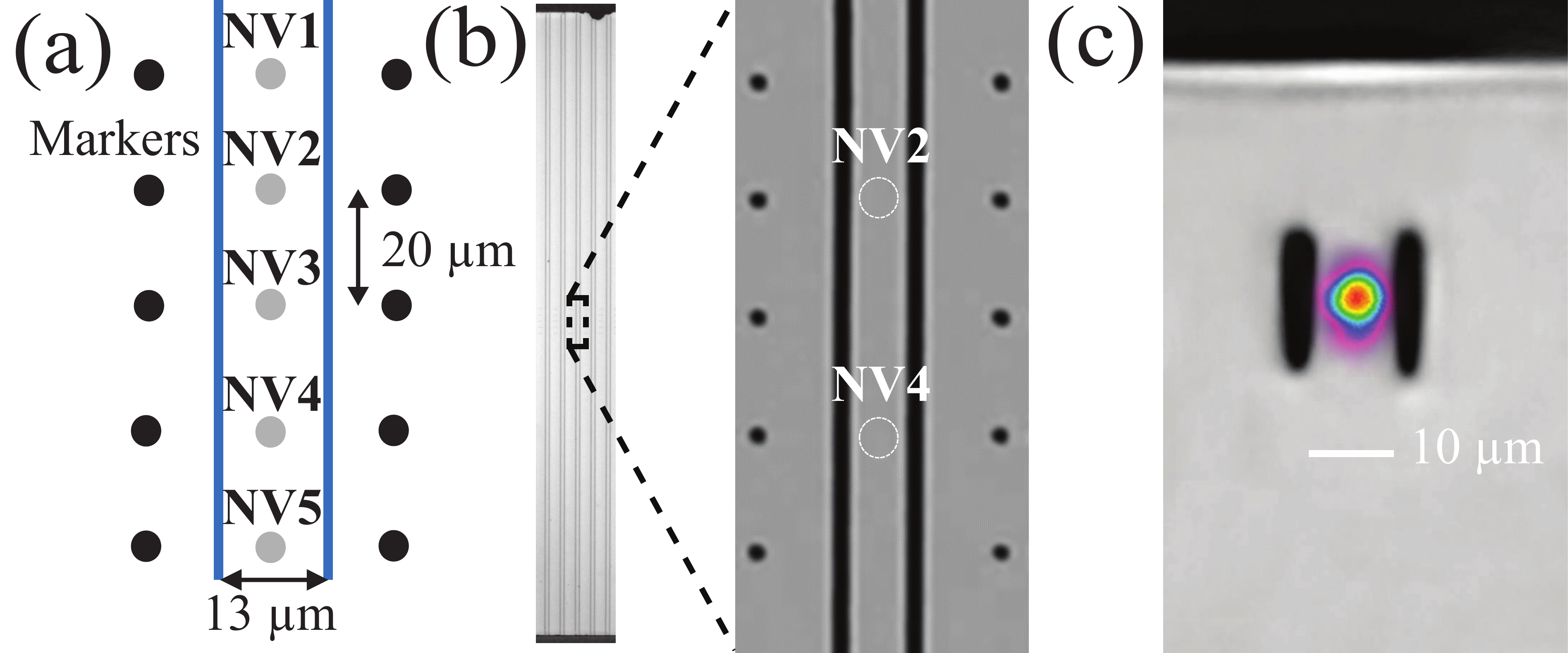}
\caption{Device overview and waveguide output mode. (a) Overhead schematic of the central \SI{80}{\micro\meter} of the waveguide-NV device. (b) Overhead optical microscope image of the full \SI{2}{\milli\meter} waveguide, with zoom-in of the central \SI{80}{\micro\meter}. (c) End view optical microscope image of type II waveguide facet with overlaid \SI{635}{\nano\meter} mode profile.}
\label{fig:Figure1a_Sample}
\label{fig:Figure1b_Sample}
\label{fig:Figure1c_Sample}
\end{figure}

\begin{figure}
\centering
\includegraphics[width=\linewidth]{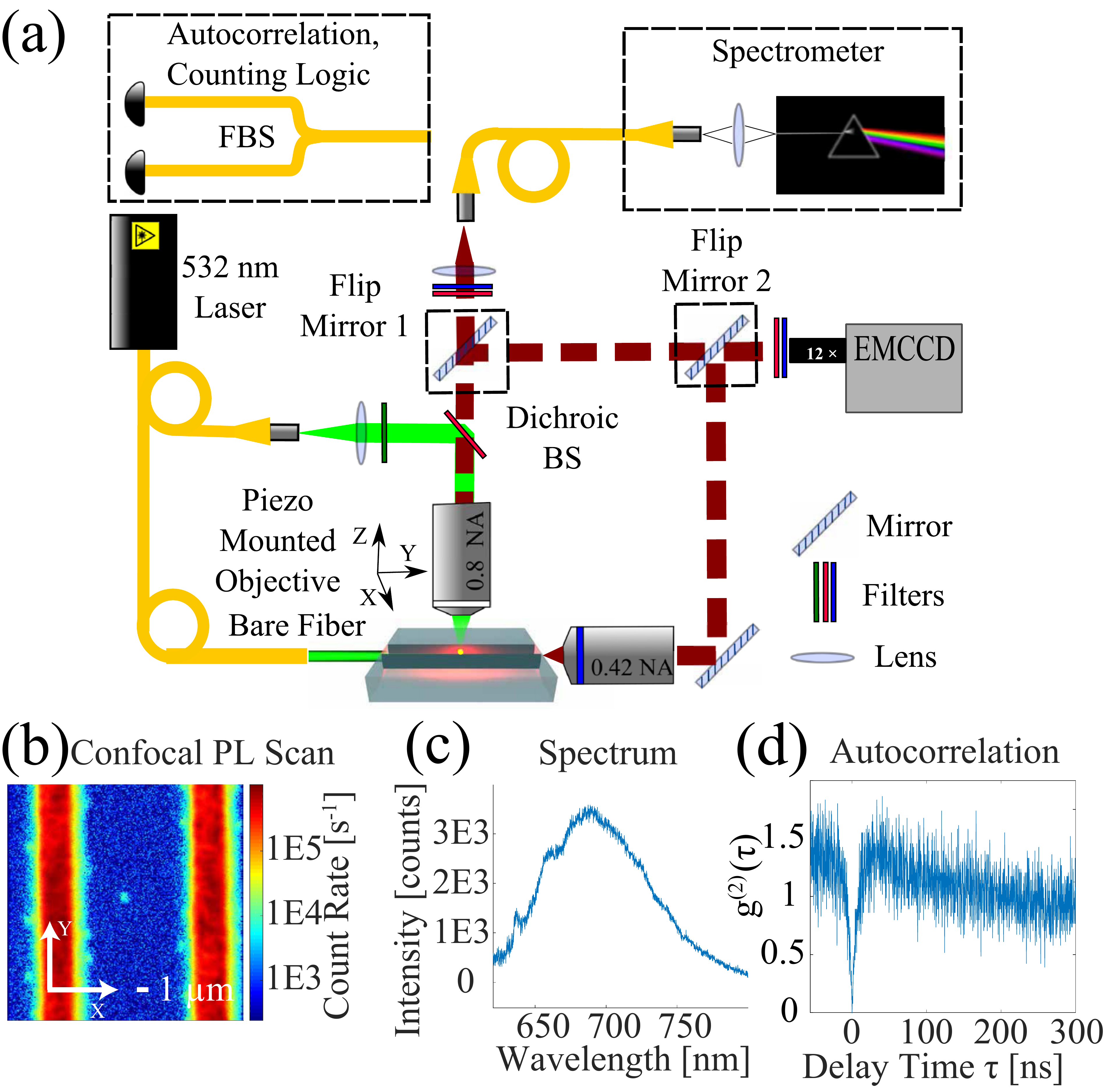}
\caption{Overhead confocal characterization. (a) Schematic of experimental setup.  (b) PL intensity map showing NV2 within waveguide. (c) PL spectrum showing NV ZPL at \SI{637}{\nano\meter} and PSB at longer wavelengths. (d) NV intensity autocorrelation revealing single photon emission.}
\label{fig:Figure2a_Setup}
\label{fig:Figure2b_XYScan}
\label{fig:Figure2c_Spectrum}
\label{fig:Figure2d_Autocorrelation}
\end{figure}

Initial overhead photoluminescence (PL) characterization of the laser-written NV centers and their interaction with the waveguide was performed using a homebuilt confocal microscope (as shown in Fig. 2(a), and described in detail in Supplementary Material \cite{SupportingInfo}).  The setup allows for \SI{532}{\nano\meter} laser excitation from overhead, focused through the microscope objective or through a butt-coupled single mode fiber coupled to the waveguide, while emitted light which is separated from the excitation using a dichroic beam splitter and interference filters can be detected and analyzed using single photon counting detectors, a spectrometer or a back-illuminated EMCCD from overhead or from the other end of the waveguide.  An overhead PL confocal scan of the NV2 trial region (positioned as shown in Fig. 2(b)) shows a bright spot corresponding to an NV center between the two bright laser modified lines. Two of the five static exposures in this study (NV2 and NV4) displayed such bright spots after annealing. The single NV formation probability at \SI{28}{\nano\joule} pulse energy including additional trials in other waveguides was \SI{31\pm 9}{\percent} which is consistent with previous work \cite{Chen2016}.  

The alignment of the NV centers close to the center of the optical waveguide is important in order for them to be well coupled. We estimate that both NV2 and NV4 are \SI{\sim 1}{\micro\meter} shallower than the center of laser-modified lines based on the confocal signal and hence the optical mode. We selected NV2 for waveguide coupling experiments as its lateral position (shifted by \SI{0.7 \pm 0.3}{\micro\meter} from the center of the waveguide, shown in Fig. 2(b)) is better centered compared to NV4 (shifted by \SI{0.9 \pm 0.3}{\micro\meter}). The \SI{\sim 1}{\micro\meter} shift of the NV2 and NV4 is larger than the maximum shifts of \SI{\sim 600}{\nano\meter} reported previously \cite{Chen2016}, which we attribute to the larger focal volume provided by our femtosecond beam delivery setup. All measurements reported below refer to NV2. 

Figure 2(c) shows the PL spectrum recorded from the NV with the \SI{650}{\nano\meter} long-pass filter removed, where the characteristic zero phonon line (ZPL) at \SI{637}{\nano\meter} and the broadband phonon side band (PSB) are clearly visible. An intensity autocorrelation measurement $g^{(2)}(\tau)$ of this is shown in Fig. 2(d), with $g^{(2)}(0)$ well below 0.5, characteristic of single photon emission.

In order to demonstrate the ability to guide \SI{532}{\nano\meter} excitation light through the waveguide to the NV center, the output from the excitation laser was coupled to the laser-written waveguide by butt-coupling a bare single mode fiber (460HP, Thorlabs) to the input facet of the diamond waveguide as shown in Fig. 2(a). Index matching oil was used to improve the transmission and coupling stability. A fraction of the light within the waveguide undergoes Raman scattering, with the first order Raman scattered light shifted to \SI{572}{\nano\meter}. Some of this scattered light is emitted out of the waveguide, and can be detected through the confocal microscope's collection arm with the \SI{650}{\nano\meter} long-pass filter removed in order to image the guiding properties of the waveguide. Figure 3(a) shows an overhead collection scan of the waveguide, revealing the Raman scattering due to the waveguided excitation light. The waveguide alignment was optimized by monitoring the output intensity at the output facet of the waveguide using a 0.42 NA objective (50$\times$ Plan Apo SL Infinity Corrected, Mitutoyo) coupled to a CCD.

\begin{figure}[bht]
\centering
\includegraphics[width=\linewidth]{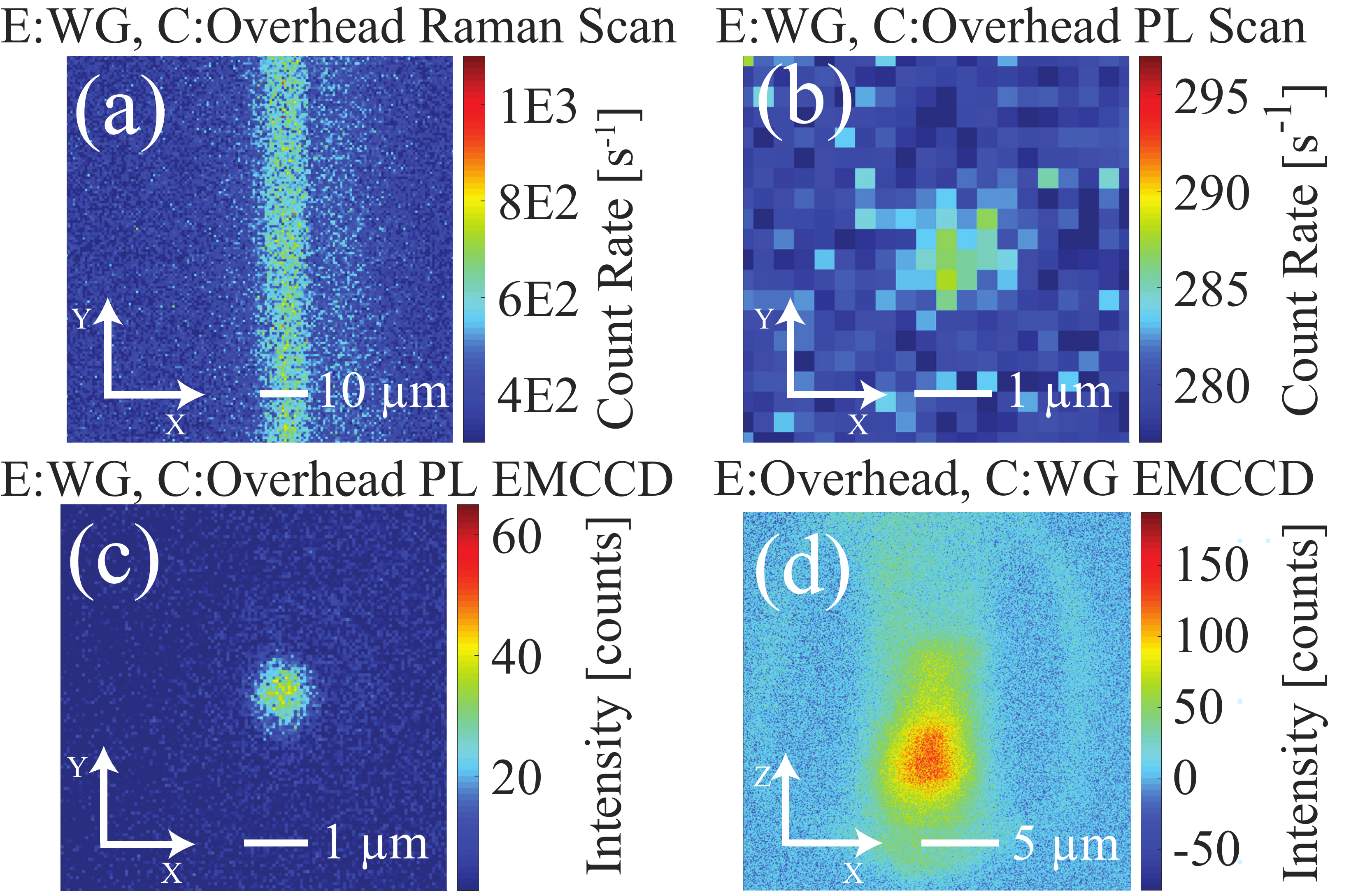}
\caption{Waveguide excitation and collection of Raman and NV fluorescence. (Titles E: Excitation mode, C: Collection mode) (a)  Waveguide excited overhead scan of Raman scattered light. (b) Waveguide excited overhead PL scan of NV2. (c) Waveguide excited overhead PL EMCCD image of NV2. (d) Overhead excited EMCCD intensity image of the of the guided NV emission at the output facet of the optical waveguide.}
\label{fig:Figure3_WGExcitationCollection}
\end{figure}

Waveguide coupled excitation of the NV center is demonstrated in Fig. 3(b) and (c), with collection of the NV center's emitted light through the overhead confocal microscope. An overhead confocal collection scan of a small area around NV2 with  waveguide excitation shows fluorescence from the expected NV position (Fig. 3(b)). Verification is achieved by imaging the  NV fluorescence with an EMCCD, which again shows emission located at the NV position (Fig. 3(c)). The relatively modest detected photon count rate of \SI{\sim 10}{counts/\second}, above the background in Fig. 3(b) is expected given that the mode size of the waveguide (\SI{\sim 5.8}{\micro\meter} FWHM) is much larger than the focused spot from free space excitation 0.8 NA lens, (\SI{\sim 0.6}{\micro\meter}) resulting in a \SI{\sim 0.01}{} theoretically estimated reduction in power density at the NV center (as compared to a 0.005 experimental estimate), in addition to coupling or waveguide losses before it reaches the NV center. Comparable count rates to the confocal microscope collection rate would be expected for similar power densities. An increase in the effective power density could be achieved by tailoring the refractive index profile of the written waveguide to achieve a smaller MFD, or more powerful excitation.

Further evidence of waveguide-NV coupling is shown in Fig. 3(d) where the NV center is excited from above through the confocal microscope's objective, while emission is collected through the optical waveguide. The waveguide output facet is imaged onto the EMCCD through a 0.42 NA objective and using flip mirror 2 (as shown in Fig. 2(a)). The signal was filtered using a \SI{650}{\nano\meter} long-pass filter, and corrected for any remaining background by subtracting the recorded signal when exciting the waveguide near the NV center.  The signal to background ratio (excluding dark counts) is 0.7. The intensity distribution of the NV emission at the waveguide output (Fig. 3(d)) shows a similar size and shape as the output mode when coupling a \SI{635}{\nano\meter} wavelength laser diode to the waveguide (Fig. 1(c)).  A theoretical estimate of the absolute coupling efficiency to the waveguide mode relative to the total emission yields \SI{9d-4}{}.  Thus the relative collection efficiency of the waveguide compared to the microscope is \SI{\sim 0.03}{} (compared to an experimental estimate of 0.012). An increase in this coupling efficiency would be expected for smaller MFD waveguides (see Supplementary Material \cite{SupportingInfo}).

In summary, we have demonstrated a millimeter length-scale integrated quantum photonic chip in diamond, fabricated entirely with femtosecond laser writing. The optical waveguides and static exposures to produce vacancies were fabricated in a single processing step, to within micron alignment. We verified through confocal microscopy that single NVs could be formed in quantum grade diamond at depths shallower than \SI{30}{\micro\meter} without a SLM compensating for spherical aberration \cite{Chen2016}. 

Importantly, we showed that optical waveguides can be used to excite and collect light from single NVs. The relatively weak NV coupling to these \SI{\sim 10}{\micro\meter} diameter waveguides do not allow for high photon collection efficiency required for quantum applications.  However it should be possible to improve the collection efficiency substantially through optimization of the geometry of the type II waveguide to reduce the mode size, as well as through the incorporation of Bragg grating waveguides\cite{Bharadwaj2017} to reduce the group velocity or ultimately engineer an optical cavity within the waveguide.  The femtosecond laser writing technique provides a simple, rapid prototyping route to millimeter length-scale arbitrary 3D photonic structures and deterministically placed NV centers, without the need for multistage clean-room procedures.  Such structures could open the door for exciting new possibilities in quantum information processing.  Waveguide coupled NV centers are also immediately relevant to spin based sensing applications such as magnetometry, electrometry or thermometry.  Here the formation of high density NV ensembles within the waveguide should enable high sensitivities even in the regime of low photon collection efficiency \cite{Gazzano2016,Clevenson2015,Chen2017}.  The integration and stability provided by monolithic optical waveguides significantly simplifies broadband sensing applications where nanoscale resolution provided by scanning probe or laser scanning confocal microscopy is not required. The further integration of these waveguides with microfluidic channels, which could be made through the etching of graphitic tracks written with femtosecond laser pulses or ion beam implantation open up the possibility of lab on chip applications \cite{Jedrkiewicz2017}.

\section{Supplementary Material}
See supplementary material for the supporting content.  This includes details on device fabrication, NV characterization, NV creation statistics, low temperature NV spectra, and theoretical and experimental estimates of waveguide NV coupling efficiency.

\begin{acknowledgments}
The authors acknowledge support from FP7 DiamondFab CONCERT Japan project, DIAMANTE MIUR-SIR grant, FemtoDiamante Cariplo ERC reinforcement grant, University of Calgary's Eyes High Postdoctoral Fellowship program, NSERC Discovery Grant, NSERC Discovery Accelerator Supplement, NSERC Research Tools and Instruments, CFI.  We thank Prof. Guglielmo Lanzani and Dr. Luigino Criante for the use of the FemtoFab facility at CNST - IIT Milano for the laser fabrication experiments.
\end{acknowledgments}

\bibliography{WGNV}

\end{document}